# Stroke Volume Estimation by Thoracocardiography is Better When Glottis is Closed

Enas Abdulhay and Pierre Baconnier *Member, IEEE*

*Abstract*—**Thoracocardiography approach pretends to non-invasively monitor stroke volume by inductive plethysmographic recording of ventricular volume curves by a transducer placed on the chest. The purpose of this study was to investigate the potential of thoracocardiography to estimate stroke volumes while apnea with open glottis. We hypothesized that, when glottis is open, stroke volumes would be better estimated if airways flow curves were taken into account.**

## I. Introduction

RECENTLY and in the light of some publications [1], [2],[3], it has seemed possible to develop non invasive hemo-dynamic monitoring tools for the ventilated patients.

Thoracocardiography (TCG) is supposed to be a non-invasive continuous monitoring of stroke volume and cardiac mechanical performance by inductive plethysmography. In this approach, respiratory waveforms [4], [5], [6] are measured by an inductive plethysmography transducer placed transversely around the chest at a level near the xiphoid process. Left ventricular volume could be extracted by ECG triggered ensemble averaging and digital band-pass filtering of respiratory waveform in order to suppress low frequency harmonics, related to respiration and other body movements, as well as high frequency electrical noise.

Heart mechanical activity induces two types of non-invasive observable volume variations :

1) the volume variations observable on inductive plethysmography signal and taken into account by thoracocardiography approach [4], [6] .
2) the volume variations observable on pneumotachogram (PNT) and called cardiogenic oscillations [7], [8].

In the first type, the heart being an element of thoracic cavity content, cardiac volume variations cause rib cage volume variations. For instance, during cardiac contraction, the left heart expels its content mainly out of the thoracic cavity and its volume fall causes a decline of pressure within the thorax. This causes a decrease of the rib cage volume. Generally, a varying intra-thoracic blood volume induces a varying thoracic volume.

In the second type, during breathing, small fluctuations were found in the flow signal derived from trunc plethysmography of five newborn infants at the same rate as the heart rate [7]. These cardiac related changes in airflow amounted to approximately one sixth of the tidal airflow. During apnea with an open glottis, Dahlstrohm et al. [9] showed in adults that there was an outward flow with every cardiac diastole and an inward flow with every systole. This was attributed to the cyclic changes of the intra-thoracic blood volume.

Putting these data together suggests that in apnea with open glottis, heart volume changes are distributed between a varying thorax volume and a varying air volume passing through airways opening.

In the present work, we investigate the potential bias on stroke volume estimation with TCG approach induced by neglecting cardiogenic oscillations. Thus, we recorded inductive plethysmography signal (RIP) in two cases: *first*, apnea with closed glottis, *second*, apnea with open glottis and compared TCG stoke volumes. We also recorded PNT signal during apnea with open glottis, measured stroke volume as seen by pneumotachography and further calculated the sum of stroke volumes of corresponding cycles on filtered pneumotachography and RIP. We used a simple model of thoracic mechanics to explain our results. This model suggests the sum of stroke volumes to provide a more accurate heart stroke volume estimation.

## II. Materials and Methods

### A. Subjects and Protocol

*5* trained *(2 males and 3 females)* healthy seated volunteers participated in the study (Table. I). Before recording, they were given 10 minutes for relaxation. All study participants provided informed consent. The study was approved by the relevant ethics committee (CHU Grenoble).

Subjects were asked to make 20 voluntary end expiratory prolonged apneas (min=3 sec, max=28 sec, mean=11 sec), 10 with open glottis and 10 with closed glottis. Volunteers were asked not to move between the first and the second recording.

TABLE. I.
SUBJECTS CHARACTERISTICS

| Subject | Age | Sex | Height (m) | Weight (kg) |
|---|---|---|---|---|
| 1 | 57 | M | 1.89 | 89 |
| 2 | 26 | M | 1.65 | 58 |
| 3 | 28 | F | 1.76 | 63 |
| 4 | 30 | F | 1.52 | 75 |
| 5 | 25 | F | 1.65 | 55 |

Manuscript received April 16, 2007. Enas Abdulhay is with the PRETA Team (Physiologie Respiratoire Expérimentale Théorique et Appliquée), TIMC, Faculté de Médecine, 38706 La Tronche cedex, France (phone: 0033--456520057; fax: 0033-456520033; e-mail: Enas.Abdulhay@ imag.fr).

Pierre Baconnier, is with the PRETA Team (Physiologie Respiratoire Expérimentale Théorique et Appliquée), TIMC, Faculté de Médecine, 38706 La Tronche cedex, France (phone: 0033-476765047; fax: 0033-476768844; e-mail: Pierre.Baconnier@ imag.fr).

## B. Materials

Rib cage and abdomen cross sectional area changes were recorded with a computer-assisted respiratory inductance plethysmography vest (Visuresp, RBI) installed 10 minutes before recording. Electrocardiogram was also recorded. During apnea with open glottis, subjects wore a face mask on which was mounted a calibrated flowmeter (Fleish head No.1) connected to a differential pressure transducer (163PC01D36, Micro Switch), turned on 30 minutes before recording. A nose clip was used during apnea with closed glottis.

All outputs were connected to an A/D converter connected to a computer. Thorax, Abdomen, Flow, and ECG signals were digitized at a rate of 100 Hz.

## C. Analysis

The method used in [10] was applied to obtain a calibrated respiratory inductance plethysmography (RIP) volume signal (VRIP,og) and (VRIP,cg) corresponding to open and closed glottis apneas, respectively. Alveolar air volume variations (VPNT) were calculated as the integral of the flow measured by pneumotachography. A band pass filter [0.7*(cardiac frequency) Hz – 10 Hz] was applied to (VRIP) according to [4], [5], [6] and to the integrated flow signal (VPNT) in order to obtain band pass filtered RIP and PNT volume signals (VRC and VA, respectively)

Beat to beat stroke volumes were measured as the amplitude over a cardiac cycle of these signals and were called differently when they were obtained during an apnea with open (SVRC,og), or closed (SVRC,cg) glottis.

## D. Model

We developed a simple trunc model to simulate cardiac mechanical activity, rib cage internal mechanics, airways resistance to flow during apnea with open and closed glottis as well as during respiration. This model (see annex) allowed us to get a better understanding of the aforementioned types of non-invasively observable volume variations induced by heart mechanical activity.

The basic idea of this model consists in standing that at any time the heart volume change distributes between rib cage and airways opening volume changes. Heart stroke volume is then better estimated by the sum of stroke volumes of rib cage and airways opening. Thus, we also measured stroke volume seen by pneumotachography (SVA) and calculated this sum (SVsum = SVRC,og + SVA)

# III. Results

Band pass filtered RIP volume signal obtained during an apnea with closed glottis (VRC,cg, subject # 3) is presented in figure 1 together with electrocardiogram. Figure 2 presents signals obtained during an open glottis apnea (VRC,og, VA) on the same subject. On this recording it appears that (VRIP) and (VPNT) do not vary in parallel and this is true on all open glottis recordings.

Results of stroke volumes comparisons are presented in table II. Differences between mean (SVRC,og) and (SVRC,cg) were significant for all subjects ($p < 0.05$); and mean closed glottis (SVRC,cg) exceeded open glottis stroke volumes SVRC,og by 18% ± 3% of SVRC,cg .

Differences between mean [SVRC,og + SVA] called (SVsum) and (SVRC,cg) were non-significant for all subjects except subject 4 ; and mean cardiogenic oscillations stroke volumes during open glottis apnea (SVA) amounted to approximately 20% ± 2% of mean (SVRC,cg).

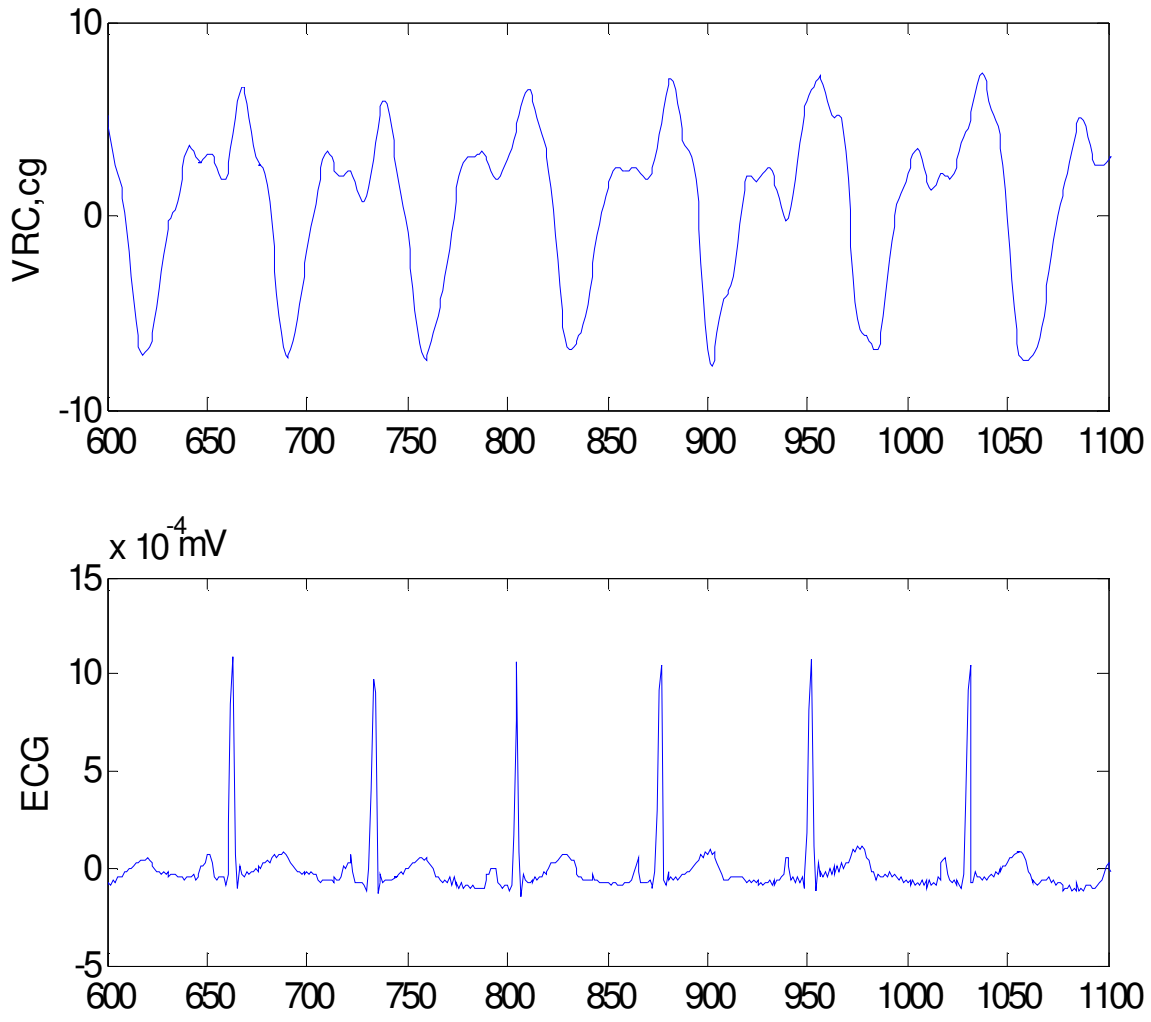

*Fig. 1. VRC,cg (in arbitrary units) and electrocardiogram (in mV) vs. points index, obtained during closed glottis apnea for subject 3. Values on the first y-axis are the real values*1000.*

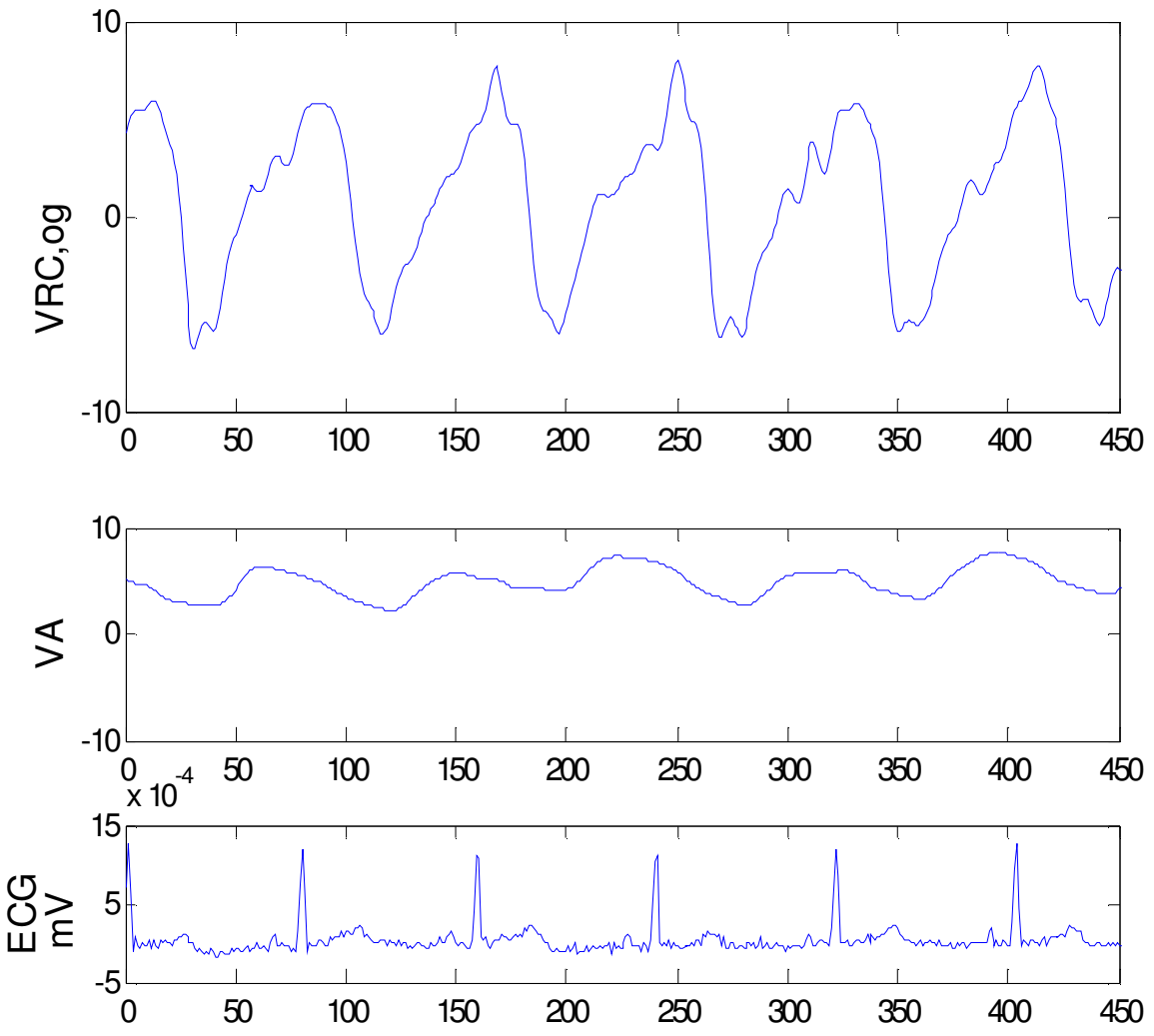

*Fig. 2. VRC,og, VA (in arbitrary units) and electrocardiogram signals obtained during open glottis apnea vs. points index for subject 3. Values on the first and second y-axis are the real values*1000.*

TABLE II
STROKE VOLUMES (IN ARBITRARY UNITS) COMPARISON FOR ALL SUBJECTS

| # | $SV_{RC,og}$ Mean (*SD*) | *p* og *vs* cg | $SV_{RC,cg}$ Mean (*SD*) | *p* cg *vs* sum | $SV_{sum}$ Mean (*SD*) |
|---|---|---|---|---|---|
| 1 | 0.020 (0.005) | **0.003** | 0.024 (0.005) | **NS** | 0.025 (0.005) |
| 2 | 0.009 (0.002) | **0.013** | 0.011 (0.002) | **NS** | 0.011 (0.002) |
| 3 | 0.011 (0.003) | **0.009** | 0.013 (0.002) | **NS** | 0.014 (0.003) |
| 4 | 0.010 (0.002) | **0.004** | 0.012 (0.002) | **0.02** | 0.012 (0.003) |
| 5 | 0.003 (0.001) | **<0.0001** | 0.004 (0.001) | **NS** | 0.004 (0.002) |

* *p :Statistical p value obtained from student T test.*

## IV. Discussion

The thoracocardiography rule supposes that the origin of cardiac signal is the left ventricle if mean expiratory exceeds mean inspiratory stroke volumes during spontaneous breathing. We compared $V_{RC}$ stroke volumes during respiration and we obtained the same result as thoracocardiography, which indicated that our results obtained with a classical respiratory inductive plethysmography (abdominal and thoracic coils) are coherent with those obtained with a one coil system.

Cardiac cavities are subjected to intrathoracic pressure and volume variations caused by breathing. On the other hand, ventricular blood volume has an influence on intrathoracic pressure and volume. The fact that right ventricle is, *first*, connected upstream to the lung, *second*, has a common wall (the interventricular septum) with the left ventricle and, *third,* ventricular blood volume varies with an equal amplitude to the stroke volume may induce a fall followed by a rise of intra-pulmonary pressure with every heart beat. Intra-pulmonary pressure rise may follow heart filling till the ejection onset ($T_{ej}$) whereas intra-pulmonary pressure fall follows ejection. As Alveolar air volume varies correspondingly to intra-pulmonary pressure, an inward air flow following systole and an outward air flow following diastole are expected.

During closed glottis apnea, thoracocardiography records the whole outward movement caused by heart filling because air is trapped inside the rib cage. On the other hand, during open glottis apnea, although an outward flow occurs, it is not taken into account.

Our model (see appendix) predicts evidently a higher stroke volume during closed glottis than during open glottis apnea (see figure 3).

Moreover, our model explains the fact that alveolar and rib cage volume variations are not in phase, as evidenced by the loops of $V_{RC,og}$ vs $V_A$ (figure 4).

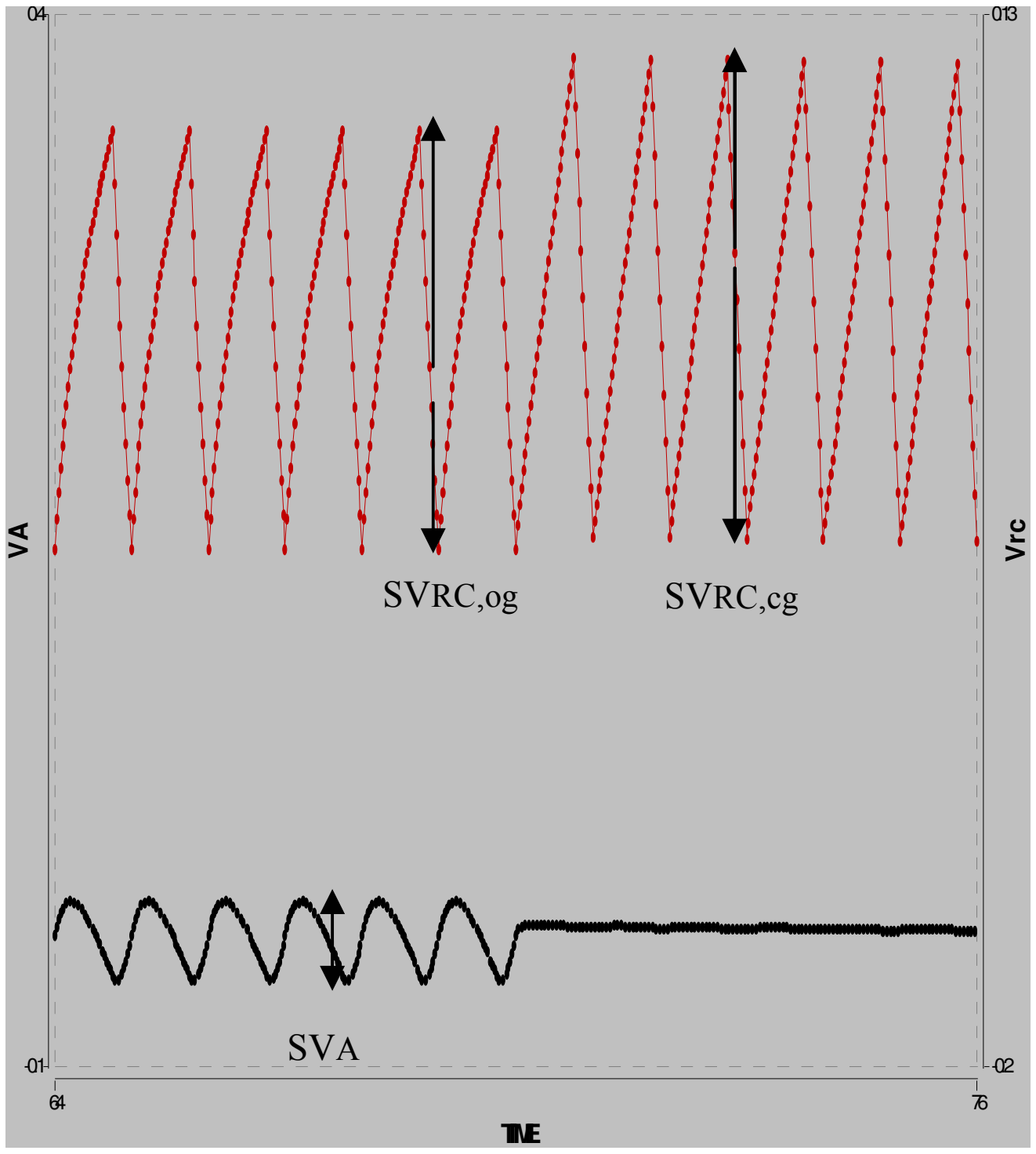

Fig. 3. *Simulation of an apnea with open glottis (6 cardiac cycles) followed by apnea with closed glottis (6 cardiac cycles). Curve at the top represents $V_{RC}$ (L) vs time and at the bottom $V_A$ (L) vs time. $SV_{RC,cg}$ is higher than $SV_{RC,og}$*

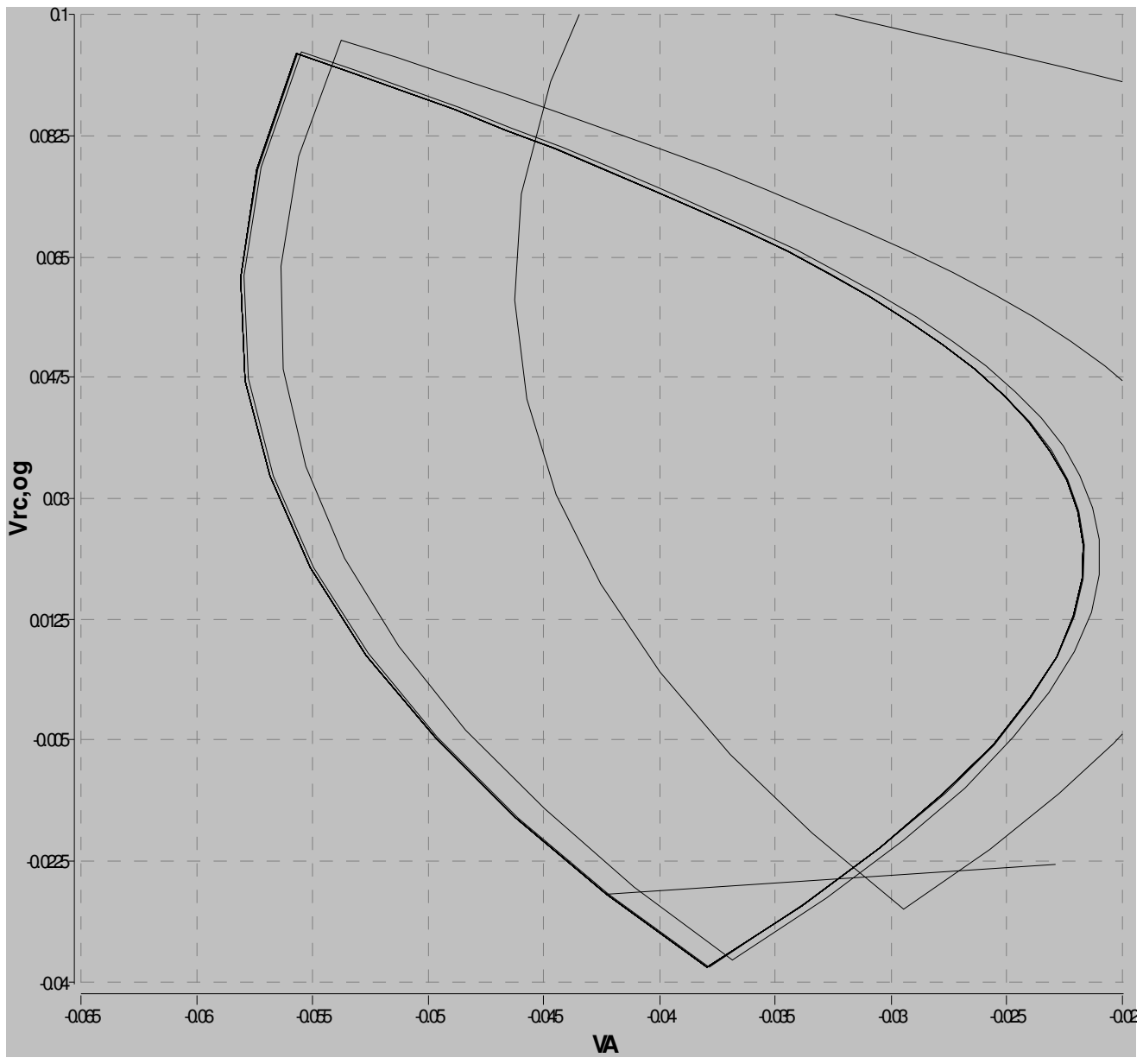

Fig. 4. *$V_{RC,og}$ vs. $V_A$ as predicted by model.*

This feature is similar to that observed in subject #3: when signals presented in figure 2 are presented in a x-y mode, one can observe loops which indicate that these signals are not in phase (see figure 5).

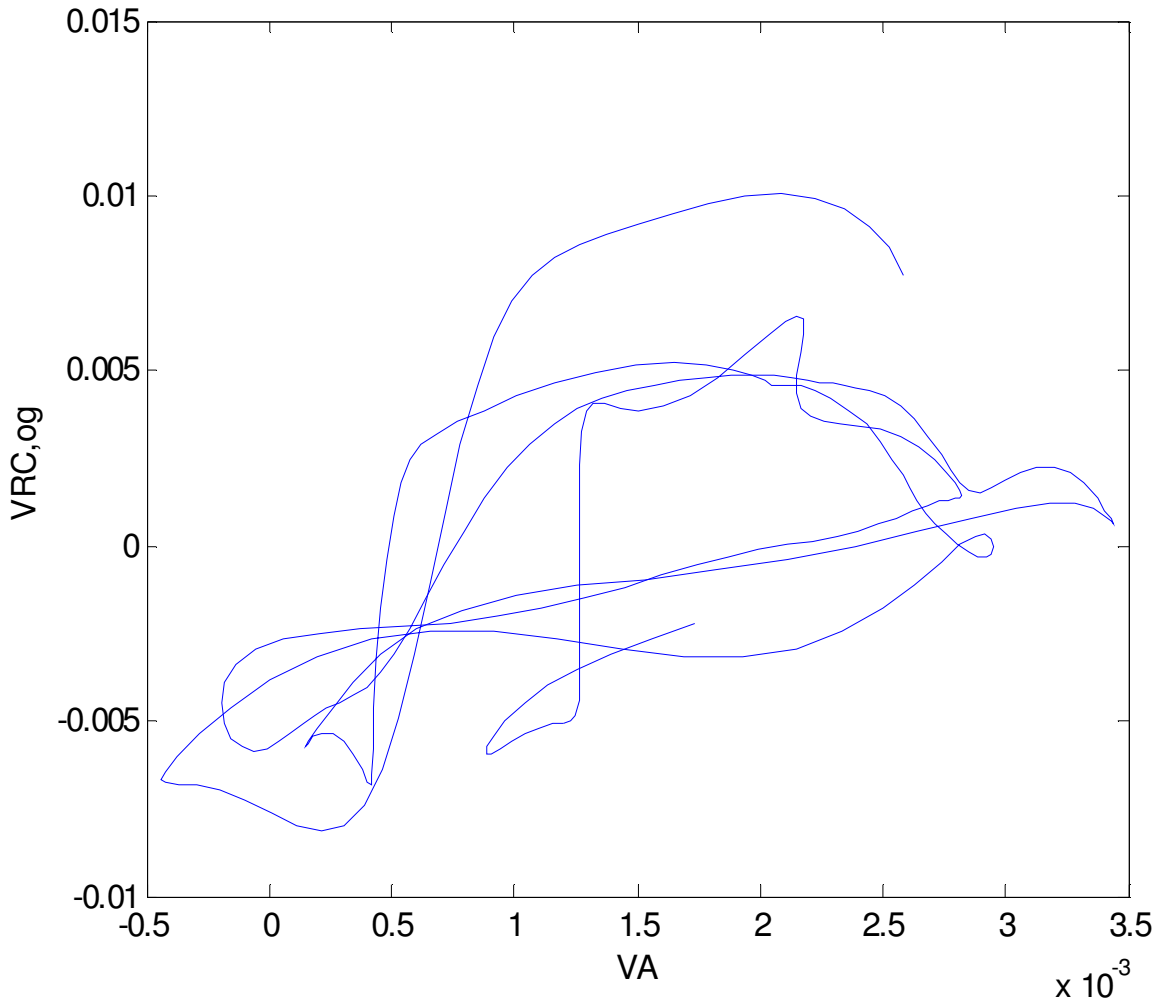

*Fig. 5. $V_{RC,og}$ vs. $V_A$ from signals of subject 3 during an open glottis apnea.*

## V. CONCLUSION

Our results demonstrate that stroke volume measured on RIP signal recorded during open glottis apneas differs from stroke volume obtained from closed glottis conditions. Moreover, in order to test the hypothesis that the sum of filtered RIP and PNT signals stroke volumes provides a better estimation of heart stroke volume, we compared $SV_{RC,cg}$ with both $SV_{RC,og}$ and $[SV_{RC,og} + SV_A]$. The comparison of results shows that $[SV_{RC,og} + SV_A]$ is more accurate than $SV_{RC,og}$ alone because the sum does not neglect air flow induced by heart beating.

## APPENDIX

Our model consists in three interconnected elements: rib cage, heart and lung. The relationship between rib cage, alveolar and intra-thoracic blood volumes (respectively $V_{RC}$, $V_A$ and $V_h$) can be presented by the following equation:

$$V_{RC} = V_h + V_A \qquad (1)$$

The behaviour of each element is the result of specific mechanism.

1) *Cardiac mechanical activity*: this activity is represented by the periodic (cardiac frequency, $F_c$) changes of intra-thoracic blood volume. In every cardiac cycle, intra-thoracic blood volume varies with an amplitude equal to the stroke volume. In a heart cycle (cardiac period $T_h = 1/F_c$) there are two phases: filling till ejection onset ( $T_{ej}$ we arbitrarily choosed $T_{ej} = T_h$*3/4) and ejection. Between 0 and $T_{ej}$, heart volume ($V_h$) increases linearly with time from 0 to stroke volume ($V_{str}$). Between $T_{ej}$ and $T_h$, heart volume decreases linearly with time from $V_{str}$ to 0.

2) *Rib cage elasticity*: in the absence of respiratory muscle activity, the chest wall is represented by a purely elastic compartment (elastance $E_{cw}$). The rib cage volume ($V_{RC}$) is driven by pleural pressure($P_{pl}$):

$$P_{pl} = E_{cw} * V_{RC} \qquad (2)$$

3) *Lung behavior*: the lung is simulated by an elastic compartment (elastance $E_l$) connected to the atmosphere by a resistive tube (resistance $R_{aw}$) submitted to pleural pressure. Its behaviour follows the equation:

$$- P_{pl} = E_l * V_A + R_{aw} * dV_A/dt \qquad (3)$$

We simulated both open and closed glottis apneas with the same model by simply modifying the upper airways resistance ($R_{uaw}$):

- open glottis, $R_{uaw}$ = 1 cmH20/l/s (normal value),
- closed glottis, $R_{uaw}$ = 100 cmH20/l/s (quasi infinite value).

The other parameter values are:

$F_c$ = 1 Hz
$V_{str}$ = 0.15 L
$E_{cw}$ = 5 cmH20/L
$E_l$ = 5 cmH20/L